\documentclass[5p,number,sort&compress,times,10pt,a4paper,twoside]{elsarticle}
\usepackage{graphicx}
\usepackage[breaklinks=true]{hyperref}
\usepackage{breakurl}

\begin{document}

\title{Zero-field $^{29}$Si nuclear magnetic resonance signature of helimagnons in MnSi}
\author[cea]{P. Dalmas de R\'eotier}
\author[cea]{A. Yaouanc}

\address[cea]{Universit\'e Grenoble Alpes, CEA, Grenoble INP, IRIG-PHELIQS, F-38000 Grenoble, France}

\begin{abstract}
  The low temperature dependence of the nuclear magnetic resonance frequency and spin-lattice relaxation rate measured in the chiral magnet MnSi by Yasuoka and coworkers [J.\ Phys.\ Soc.\ Jpn.\ {\bf 85}, 073701 (2016)] is interpreted in terms of helimagnon excitations. The theoretically predicted gapless and anisotropic dispersion relation which is probed at extremely small energy is experimentally confirmed. Whenever comparison is possible, the results are found quantitatively consistent with those of the inelastic neutron scattering and muon spin rotation and relaxation techniques. Further studies are suggested.
\end{abstract}


\maketitle

\section{Introduction}
\label{Intro}

The metallic cubic compound MnSi (space group $P2_13$) has a long history of attracting the attention of physicists. It displays a chiral magnetic structure below $T_c \approx 29.5$~K in zero external magnetic field \cite{Ishikawa76}, characterized by a very small propagation wavevector ${\bf k}$ parallel to a three-fold axis:\footnote{The modulus of the propagation wavevector is about $k = 0.35 \, {\rm nm}^{-1}$ in the low temperature range of interest here \cite{Ishikawa76,Grigoriev06,Janoschek13}.} the pitch of the structure is $\simeq$ 18~nm. It is a prototypal weak itinerant ferromagnet \cite{Moriya85} and it hosts a skyrmion lattice phase \cite{Muhlbauer09b}.

Spin waves are the primary low-energy collective spin excitations observed in magnetic materials. They are traditionally studied by inelastic neutron scattering experiments. These measurements are generally aimed to the determination of the spin waves dispersion relation from which the magnetic interactions can be quantified. In helimagnets, the quasiparticles associated with the spin waves are known as helimagnons. As a matter of fact, owing to the long wavelength of the helical modulation of MnSi, the study of these excitations has been essentially restricted to wavevectors $q \gtrsim k$ in line with the momentum and energy resolution range accessible to inelastic neutron scattering spectrometers \cite{Ishikawa77,Semadeni99}. The attention has recently focused on the formation of bands of helimagnons \cite{Janoschek10,Kugler15,Schwarze15}. 

Because of the specificity of a long-pitch helical magnetic structure, the study of the dispersion relation for wavevectors $0< q \lesssim k$, i.e.\ very near the first Brillouin zone center, is of special interest. From the universal behavior of magnets \cite{Kittel63}, the helimagnon energy $\hbar\omega({\bf q})$ is expected to be linear or quadratic in $q$, depending whether ${\bf q}$ is parallel or perpendicular to ${\bf k}$. Indeed, in a plane perpendicular to ${\bf k}$, the spins are ferromagnetically arranged, whereas the structure along the direction of ${\bf k}$ is  antiferromagnetic with a long periodicity.

Given the magnetic interaction energy range at play, probing the wavevector region $0< q \lesssim k$ requires an experimental technique with a sensitivity to excitations in the microelectronvolt energy range. Nuclear magnetic resonance (NMR) is such a technique which was recognized for its ability to probe extremely low energy magnons a long time ago; see e.g.\ Ref.~\cite{Jaccarino57}. Here we consider a recent NMR study of MnSi in its magnetically ordered phase \cite{Yasuoka16}. We restrict ourselves to data recorded in zero applied field \cite{Gossard59,Riedi89}: the MnSi magnetic structure is then purely helical and the dispersion relation is not affected by any additional term. We show that the NMR spectra can be consistently interpreted in terms of gapless helimagnons relevant to the free energy written by Bak and Jensen \cite{Bak80} or Nakanishi {\em et al} \cite{Nakanishi80}, with physical parameters in good agreement with previous estimates.

The organization of the paper is as follows. We present the experimental NMR data in section \ref{Data} and the model used for their interpretation in section \ref{Model}. Section \ref{Discussion} offers a discussion of the results. A summary and a conclusion are given in section \ref{conclusion}. Finally an appendix provides mathematical details for the model.

\section{Experimental results}
\label{Data}

Several zero-field NMR measurements have been performed on MnSi; see e.g.\ Refs.~\cite{Motoya76,Motoya78,Thessieu98,Yu04}. Here we will consider the data from Ref.~\cite{Yasuoka16} for which a temperature dependence is available. This is $^{29}$Si NMR, where the spin of the resonant nucleus is 1/2. No electric quadrupole interaction is present and the nuclear energy levels are therefore only subject to Zeeman splitting. The data have been recorded on powder samples.

\begin{figure}
 \includegraphics[width=\linewidth]{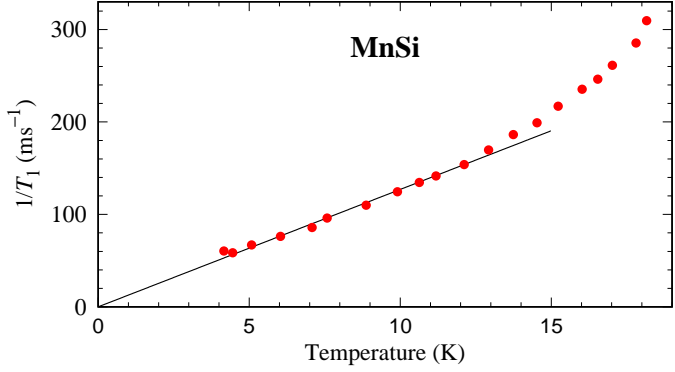}
 \caption{Temperature dependence of the inverse of the zero-field $^{29}$Si NMR spin lattice relaxation time $T_1$. The data are reproduced from \cite{Yasuoka16}. The full line results from the linear fit of $1/T_1$ = $b_{\rm NMR} T$ to the data recorded below 13~K with $b_{\rm NMR} = 12.7\,(1)$\,ms$^{-1}$K$^{-1}$. Note that $b_{\rm NMR}$ is fully consistent with the value 12.5~ms$^{-1}$K$^{-1}$ found in Ref.~\cite{Thessieu98}.}
\label{rate}
\end{figure}

\begin{figure}
  \includegraphics[width=\linewidth]{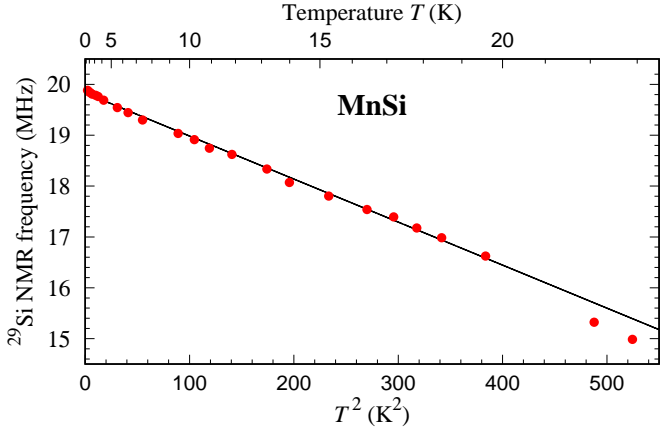}
  \caption{Zero-field $^{29}$Si NMR frequency $\nu$ versus the square of the temperature. The data are taken from Ref.~\cite{Yasuoka16}. The full line represents a fit of $\nu(T) = \nu(0) (1 - T^2/T_{\rm he}^2 )$ to the data below 20~K, with $\nu(0)$ = 19.83\,(2)\,MHz and $T_{\rm he} = 48.4\,(3)$\,K.}
 \label{frequency}
\end{figure}

In Fig.~\ref{rate}, we display the spin lattice relaxation rate $1/T_1$ versus the temperature $T$. This rate is found proportional to the temperature up to nearly 15~K, i.e.\ for $T < T_c/2$.\footnote{The $1/T_1 \propto T$ behavior is reminiscent of the relaxation observed in free electrons metals \cite{Korringa50}. This process does not apply for strongly correlated electron systems such as MnSi \cite{Corti07,Yasuoka16}.}  Figure \ref{frequency} shows the temperature dependence of the center frequency $\nu$ of the $^{29}$Si resonance. This quantity is a good and accurate measure of the temperature dependence of the staggered magnetic moment \cite{Yasuoka16}. From Fig.~\ref{frequency} we conclude that the Mn magnetic moment decreases as the square of the temperature up to $\approx 2 T_c/3$.


\section{Model}
\label{Model}

A spin-lattice relaxation process in a magnet involves a nuclear spin transition which is associated with the absorption or emission of spin waves \cite{Moriya56a}. In general a direct process with a single spin wave is not allowed because of the energy conservation requirement. Indeed, the Zeeman energy of the nuclear probe is usually much lower than the minimum energy of a spin wave.
Conversely energy conservation can be satisfied by processes involving the simultaneous creation and annihilation of spin waves \cite{Moriya56a,vanKranendonk56}, in particular Raman spin wave processes. These processes lead to a spin lattice relaxation rate varying strongly with temperature, at least as $T^2$ \cite{Mitchell57}, at odds from the observation in MnSi.

For an insight into the variation of $1/T_1$ we recast to the expression of this quantity. The magnetic interactions between the nuclear and the electron spins are responsible for the spin-lattice relaxation with a rate \cite{Moriya64,Moriya56a,Moriya62}
\begin{eqnarray}
  \frac{1}{T_1} & = & \frac{^{29}\gamma^2}{2} \int_{-\infty}^{\infty} \langle\{\delta B_-(t)\delta B_+\}\rangle \exp(-i \omega_0 t)\,{\rm d} t
  \label{1/T1}
\end{eqnarray}
where $\delta B_\alpha$, $\alpha \in \{x,y,z\}$, is the fluctuating part of Cartesian component $\alpha$ of the magnetic field acting on the $^{29}$Si spin.\footnote{As usual, $\delta B_\pm \equiv \delta B_x \pm i \delta B_y$.} The quantity $^{29}\gamma$ is the gyromagnetic ratio of the $^{29}$Si nucleus and $\omega_0$ is its resonance angular frequency in the mean internal field which defines the $z$ axis. The quantity $1/T_1$ is then proportional to the time Fourier transform of the dynamical field correlation $\langle\{\delta B_-(t)\delta B_+\}\rangle$ taken at $\omega = \omega_0$. This field correlation can be expressed in terms of the electronic spin correlations $\langle\{\delta S_\alpha(\omega = \omega_0)\delta S_\beta\}\rangle$ through hyperfine coupling terms. In the temperature range of interest here, i.e.\ $T<$~15~K, the thermal expansion of MnSi is sufficiently small that these coupling terms can be considered as constant \cite{Matsunaga82,Riedi73,Edwards76}. Therefore the temperature dependence of $1/T_1$ tracks that of $\langle\{\delta S_\alpha(\omega = \omega_0)\delta S_\beta\}\rangle$.

The spin correlations $\langle\{\delta S_{x,y}(\omega = \omega_0)\delta S_{x,y}\}\rangle$ can be readily expressed in terms of magnon operators through the linear Holstein-Primakov transformation. In the high temperature limit which applies here since $k_{\rm B}T \gg \hbar \omega_0$, the magnon population factor is linearized, which leads to $\langle\{\delta S_{x,y}(\omega = \omega_0)\delta S_{x,y}\}\rangle$ $\propto k_{\rm B}T$. A linear temperature dependence of $1/T_1$ therefore indicates that the $^{29}$Si spins relax through a process involving the creation or annihilation of single magnons at energy equal to $\hbar\omega_0 \lesssim 0.1~\mu$eV, a bound given by the resonance frequency; see Fig.~\ref{frequency}.  The interpretation of $\nu(T)$ to which we turn now will confirm the role of these excitations in the NMR results. 

In order to model $\nu(T) \propto m(T)$ we follow Bloch's law methodology which assigns the low temperature decrease of the magnetization to spin wave excitations. Applied to ferromagnets, this model leads to the well-known $T^{3/2}$ law \cite{Kittel63}. Mathematically,
\begin{eqnarray}
m(T) = m(T = 0) \left(1 - \frac{1}{N}\sum_{\bf q} n_{\bf q}  \right),
\label{calcul_1}
\end{eqnarray}
where $n_{\bf q}$ = $[\exp(\hbar \omega_{\bf q}/k_{\rm B}T)-1]^{-1}$ is the boson occupation factor and $N$ is the number of magnetic atoms in the system. The magnon Goldstone mode of angular frequency $\omega_{\bf q}$ relevant for helimagnons has a dispersion relation which is anisotropic with respect to the components parallel and perpendicular to ${\bf k}$ of wavevector ${\bf q}$. In a first approximation \cite{Maleyev06,Belitz06},
\begin{eqnarray}
  \omega_{\bf q}^2 =  c_\parallel q_\parallel^2 + c_\perp q_\perp^4,
\label{Goldstone_simplified}
\end{eqnarray}
which has the Landau-Peierls form characteristics for lamellar structure \cite{Chaikin95,Garst17}. This form for the dispersion relation satisfies the naive expectation for helimagnons propagating either parallel or perpendicular to ${\bf k}$; see section \ref{Intro}. The elastic constants $c_\parallel$ and $c_\perp$ fulfil the relation \cite{Maleyev06,Belitz06}
\begin{eqnarray}
c_\perp      =  \frac{c_\parallel}{ 2 k^2},
\label{calcul_4}
\end{eqnarray}
so that the dispersion relation depends on a single free parameter.
However we should note that, strictly speaking, Eq.~\ref{Goldstone_simplified} is approximate. To prevent the destabilization of the magnetic order by the Landau-Peierls instability \cite{Garst17}, an additional term proportional to $q_\perp^2$ is added to the expression of $\omega_{\bf q}^2$ \cite{Belitz06}: 
\begin{eqnarray}
  \omega_{\bf q}^2 =  c_\parallel (q_\parallel^2 + c_{\rm ea} q_\perp^2) + c_\perp q_\perp^4.
\label{calcul_3}
\end{eqnarray}
Parameter $c_{\rm ea}$ gauges the influence of the exchange interaction anisotropy on the dispersion relation. It is expected to be only a few per cent \cite{Hu18}.

The evaluation of Eq.~\ref{calcul_1} is performed along the lines given in the Appendix. A quadratic decay for $m(T)$ is found,
\begin{eqnarray}
m(T) = m(T = 0) \left(1 - \frac{T^2}{T_{\rm he}^2} \right),
\label{app_10}
\end{eqnarray}
where we have defined 
\begin{eqnarray}
{T_{\rm he}} & = & 4\sqrt{\frac{3}{\pi v_0}} \frac{\hbar}{k_{\rm B}} (c_\parallel c_\perp)^{1/4},
\label{app_9}
\end{eqnarray}
and $v_0$ is the volume per magnetic ion.

\section{Discussion}
\label{Discussion}

The temperature dependence of the magnetic moment $m(T) \propto \nu(T)$ is well accounted for by Eq.~\ref{app_10} up to $\approx 20$~K; see Fig.~\ref{frequency}. This is a first indication that helimagnons are responsible for the zero-field NMR observations.

From the experimental value of $T_{\rm he}$ and Eq.~\ref{app_9} we estimate $c_\parallel c_\perp$ = $3.9\,(1)\,\times 10^{-9}$~m$^6$\,s$^{-4}$.\footnote{In the numerics, we take $v_0 = a_{\rm lat}^3/4$ since the cubic unit cell of lattice parameter $a_{\rm lat}$ = 4.558~\AA\ contains four Mn atoms.} As explained in the Appendix we have no information on the $c_{\rm ea}$ parameter except that it is small, as expected. The value for the product $c_\parallel c_\perp$ compares favorably with a recent muon spin rotation ($\mu$SR) estimate of $4.46\,(15)\,\times 10^{-9}$~m$^6$\,s$^{-4}$ \cite{Yaouanc20}. Using in addition Eq.~\ref{calcul_4}, from the NMR results we get $c_\parallel$ = 30.8\,(4)\,$\times 10^3$~m$^2$\,s$^{-2}$ and $c_\perp$ = 0.126\,(2)\,$\times 10^{-12}$~m$^4$\,s$^{-2}$. This value of $c_\perp$ is close to the value 0.11\,(1)\,$\times 10^{-12}$~m$^4$\,s$^{-2}$ found by inelastic neutron scattering experiments, admittedly performed at temperatures approaching $T_c$ \cite{Semadeni99,Sato16}. Now, the magnetic interaction energy relevant for helimagnets like MnSi and valid at long wavelength, is ${\mathcal E}_{\rm int} = \sum_{\bf q}  B_1 q^2  |{\bf S}_{\bf q}|^2/2  + i D {\bf q} \cdot ({{\bf S}}_{\bf q} \times {{\bf S}}_{-{\bf q}})$ where we have neglected the aforementioned anisotropic interaction \cite{Bak80,Nakanishi80}.\footnote{Note that, besides the terms proportional to $B_1$ and $D$ present in ${\mathcal E}_{\rm int}$, the magnetic free energy contains the additional term $A|{\bf S}_{\bf q}|^2/2$ \cite{Bak80,Nakanishi80}, which is irrelevant for helimagnon excitations.} From the relation $c_\parallel  =  \left (B_1k S/ \hbar \right )^2$ \cite{Maleyev06}, we derive $B_1$ = $2.7\times 10^{-40}$~J\,m$^2$. Noticing that, to a good approximation, $|D|=kB_1$, the quantity ${\mathcal E}_{\rm int}$ is entirely determined.

The helimagnon excitations described by Eq.~\ref{calcul_3} are gapless. This justifies the possibility of $^{29}$Si spin relaxation through a single excitation (section \ref{Model}) and accordingly the linear temperature dependence of $1/T_1$ is a second indication that helimagnons are at play in the NMR observations. From the inferred value for $c_\parallel$ and $c_\perp$ and the experimental value of $\nu$ we deduce that excitations with wavevectors $q\lesssim k/20$ are probed in the $^{29}$Si zero-field NMR experiments.

It is noticed that the proposed model fails to account for the temperature dependence of $1/T_1$ and $\nu$ respectively above $\approx T_c/2$ and $2T_c/3$. While several reasons can be invoked, the most obvious is the breakdown of the linear spin-wave theory used in section \ref{Model} when the temperature approaches $T_c$.

From the qualitative interpretation of the linear temperature dependence of $1/T_1$ and the quantitative interpretation of $\nu(T)$ we definitely conclude that (i) the zero-field NMR data bear the signature of helimagnons in MnSi, and (ii) the helimagnon dispersion relation predicted in Refs.~\cite{Maleyev06,Belitz06} is experimentally confirmed.

This conclusion rests in part on the temperature dependence of the zero-field staggered magnetization $m(T)$ at low temperature. Besides the NMR data used here and the results of a $\mu$SR study \cite{Yaouanc20}, published data showing a detailed and accurate temperature dependence are scarce. The traditional bulk magnetization measurements are not of direct help since at the macroscopic scale, MnSi is essentially an antiferromagnet rather than a ferromagnet. Estimates of $m(T)$ can in turn only be achieved from an extrapolation of measurements at relatively high field (see e.g.\ Ref.~\cite{Bloch75}) with inherent uncertainties in the extrapolation to zero-field.

The similar temperature behavior observed for the $^{29}$Si and muon probes call for two remarks about the latter. Muon-induced effects are sometimes evoked, which may locally alter the properties of the compound, so that the muon is not a passive probe of it. Actually the positive electric charge of the implanted muon could not be sufficiently screened at short length. It has already been shown quantitatively that it is not the case for MnSi \cite{Dalmas18}. Density functional theory computations also indicate that the Mn atoms are little affected by the muon presence \cite{Onuorah18}. The consistency of the $\mu$SR and NMR results provides further support to this conclusion.

The magnetic field at the muon position for some elemental ferromagnets such as Fe, Co and Ni does not track the saturation magnetization of the metals \cite{Nishida78}. This has been attributed to the zero-point motion of the muon \cite{Estreicher82,Yaouanc11,Onuorah19}. This quantum effect is obviously negligible for MnSi since $\nu(T)$ and $m(T)$ --- as deduced from $\mu$SR --- agree and $^{29}$Si has a negligible zero-point motion. 

The coupling between the $^{29}$Si probe and the system including its anisotropy has not been accurately determined. Its knowledge would provide valuable further information. In particular the relation between the field correlation and the Mn spin correlation functions would be expressed (section \ref{Model}), from which the gradient $b_{\rm NMR}$ of $1/T_1$ vs $T$ could be predicted (see Eq.~\ref{1/T1} and Fig.~\ref{rate}). The amplitude of the staggered magnetic moment could also be precisely inferred from $\nu$. In this respect the knowledge of the muon coupling constants \cite{Amato14} allowed for the determination of the absolute value of $m(T)$. In addition, the knowledge of the $^{29}$Si coupling constants would permit the interpretation of the shape of the field distribution at the probe and possibly observe the small phase shift between the helix associated with the Mn site characterized by a 3-fold symmetry axis parallel to ${\bf k}$ and the helix associated with the other Mn sites \cite{Dalmas16,Dalmas17}.

\section{Summary and conclusion}
\label{conclusion}

Published data \cite{Yasuoka16} of the temperature dependence of the spin-lattice relaxation rate and of the zero-field NMR frequency at low temperature have been successfully interpreted in the helimagnon framework. The form of the dispersion relation predicted by theory is verified. The derived parameters are quantitatively in accord with those found in inelastic neutron scattering and $\mu$SR. The gapless nature of the excitations is confirmed on a scale less than one microelectronvolt. From the value of the free parameter of the dispersion relation, the magnitude of the exchange interaction is derived.

For decades, the intermetallics MnSi has been the prototype for weak itinerant ferromagnets. Here the decay of its low temperature magnetic moment is explained in terms of conventional helimagnon excitations. This analysis therefore suggests MnSi to be a dual system composed of localized and itinerant magnetic electrons. Recent theoretical and experimental works support this picture \cite{Choi19,Chen20,Yaouanc20}.

The successful interpretation of the zero-field NMR data available for MnSi suggests to attempt similar measurements for other systems with dominant ferromagnetic exchange interactions and subdominant Dzyaloshinsky-Moriya interactions leading to a long-range helical order in zero external field. Cu$_2$OSeO$_3$ is such a system \cite{Bos08,Maisuradze11,Adams12} although the presence of two Cu sites might have subtle effects. This is not the case for FeGe or the solid solutions Fe$_x$Co$_{1-x}$Si, Co$_x$Mn$_{1-x}$Si, and Mn$_{1-x}$Fe$_x$Si that crystallise with the same crystal structure as MnSi and exhibits a long-range helical order \cite{Lebech89,Beille81,Beille83,Bannenberg18,Kindervater20}. Together with $^{29}$Si, the spin 1/2 $^{57}$Fe nucleus is suitable for zero-field NMR studies \cite{Robert60}.

\section*{Acknowledgement}
The authors are grateful to B. Roessli and A. Maisuradze for thoughtful comments to the manuscript.

\appendix

\section{Discussion about the evaluation of $m(T)$}
\label{calcul}

For the evaluation of Eq.~\ref{calcul_1} we switch to the continuum limit
\begin{eqnarray}
  \frac{1}{N}\sum_{\bf q} n_{\bf q} = \frac{V}{N} \int
  \frac{1}{\exp \left (\frac{\hbar \omega_{\bf q}}{k_{\rm B}T} \right ) -1}  
\frac{{\rm d}^3 {\bf q}}{(2 \pi)^3}.
\label{calcul_2}
\end{eqnarray}
Here, $V$ is the sample volume. The symmetry of the dispersion relation (Eq.~\ref{calcul_3}) suggests the use of cylindrical coordinates for the computation of the integral in \ref{calcul_2}. Setting
\begin{eqnarray}
c(T) & \equiv & \frac{\hbar}{k_{\rm B}T} \frac{c_\parallel c_{\rm ea}}{\sqrt{c_\perp}},
\label{calcul_5}
\end{eqnarray}
we obtain
\begin{eqnarray}
  \frac{1}{N}\sum_{\bf q} n_{\bf q} & = &\frac{1}{2\pi^2}\frac{v_0 k_{\rm B}^2 T^2}{\hbar^2\sqrt{c_\parallel c_\perp}} \label{calcul_6}\\ & 
  \times& \int_0^{\infty}\int_0^{\infty} \frac{\hat q_\perp \,{\rm d} \hat q_\parallel\,{\rm d} \hat q_\perp}{\exp\left(\sqrt{\hat q_\parallel^2+c(T) \hat q_\perp^2+\hat q_\perp^4}\right)-1},
\nonumber
\end{eqnarray}
where $\hat q_\parallel$ and $\hat q_\perp$ are dimensionless wavevectors and $v_0$ is the volume per manganese ion. 

\begin{figure}
 \begin{picture}(255,200)
 \put(0,0){\includegraphics[width=\linewidth]{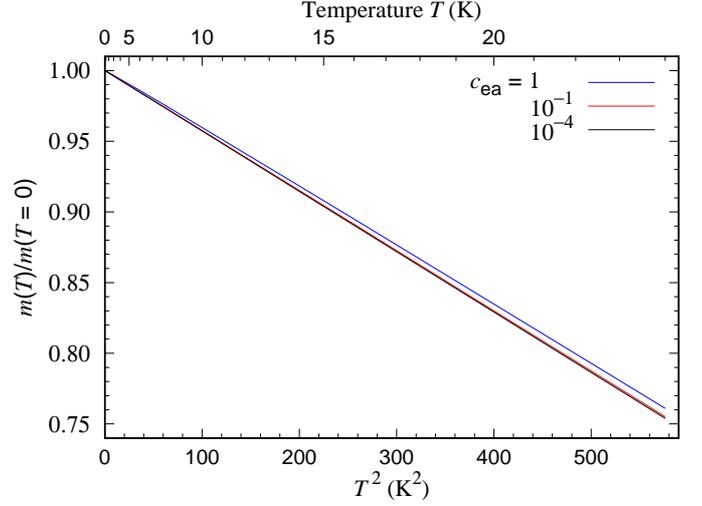}}
 \end{picture}
 \caption{Normalized magnetic moment versus squared temperature computed from Eqs.~\ref{calcul_1}, \ref{calcul_6} and \ref{calcul_5} for selected values of parameter $c_{\rm ea}$. The plot is obtained for $c_\parallel$ = $30.8\times 10^3$~m$^2$\,s$^{-2}$ and $c_\perp$ = $c_\parallel /2k^2$ = $0.126 \times 10^{-12}$~m$^4$\,s$^{-2}$, consistent with the data of Fig.~\ref{frequency}. The curves for $c_{\rm ea}$ = $10^{-1}$ and $10^{-4}$ can hardly be distinguished from each other pointing out that $m(T)$ is independent of $c_{\rm ea}$ for $c_{\rm ea} \lesssim 0.1$. Hence, $c_{\rm ea}$ cannot be determined from the analysis of $m(T)$. In all cases $m(T)$ varies linearly with $T^2$, which implies that, for the chosen values of $c_\parallel$ and $c_\perp$, the temperature dependence of $c(T)$ can be neglected even for hypothetical values of $c_{\rm ea}$ as large as 1. }
\label{m_T2}
\end{figure}

Figure \ref{m_T2} displays the temperature dependence of the magnetic moment, numerically computed from Eqs.~\ref{calcul_1}, \ref{calcul_6} and \ref{calcul_5} for different values of $c_{\rm ea}$. As can be inferred from the plot, the numerical value of the double integral in Eq.~\ref{calcul_6} is independent of $c(T)$ for any realistic value of $c_{\rm ea}$.\footnote{It is to be noted that the integral diverges for $c_{\rm ea}$ = 0, i.e.\ $c(T)$ = 0. This reflects the Landau-Peierls instability \cite{Chaikin95}. } It approaches $\pi^3/24$ for $c(T) \to 0 $. Defining parameter $T_{\rm he}$ (Eq.~\ref{app_9}), we arrive at a quadratic decay for $m(T)$ (Eq.~\ref{app_10}). This result has been first given in Ref.~\cite{Yaouanc20}.

\bibliographystyle{elsarticle-num}
\bibliography{reference}

\end{document}